\documentstyle[12pt]{article}
\newcommand{\beq}{\begin{equation}}
\newcommand{\eeq}{\end{equation}}
\newcommand{\ba}{\begin{array}}
\newcommand{\ea}{\end{array}}
\newcommand{\beqa}{\begin{eqnarray}}
\newcommand{\eeqa}{\end{eqnarray}}
\newcommand{\lsim}{\stackrel{<}{_\sim}}
\newcommand{\gsim}{\stackrel{>}{_\sim}}
\newcommand{\epoe}{\epsilon'/\epsilon}
\newcommand{\tq}{{\tilde q}}
\newcommand{\tit}{{\tilde t}}
\newcommand{\tiu}{{\tilde u}}
\newcommand{\tM}{{\widetilde M}}
\newcommand{\cA}{{\cal A}}
\newcommand{\cO}{{\cal O}}
\newcommand{\PL}[3]{{\it Phys.\ Lett.\ }        {\bf #1} {#3} {(19#2)}}

\newcommand{\PRL}[3]{{\it Phys.\ Rev.\ Lett.\ } {\bf #1} {#3} {(19#2)}}
\newcommand{\PR}[3]{{\it Phys.\ Rev.\ } {\bf #1} {#3} {(19#2)}}
\newcommand{\NP}[3]{{\it Nucl.\ Phys.\ }        {\bf #1} {#3} {(19#2)}}
\newcommand{\ZP}[3]{{\it Z.\ Phys.\ }   {\bf #1} {#3} {(19#2)}}
\newcommand{\JHEP}[3]{{\it J.\ High\ Energy\ Phys.\ }{\bf #1} {#3}
{(19#2)}}
\newcommand{\PTP}[3]{{\it Prog.\ Theor.\ Phys.\ }{\bf #1} {#3}
{(19#2)}}
\newcommand{\EPJ}[3]{{\it Eur.\ Phys.\ J.\ }{\bf #1} {#3} {(19#2)}}

\title{Precision tests and searches\\
for New Physics with $K$ decays\thanks{
Talk presented at the ``International Workshop
on $CP$ Violation in $K$'',
18-19 December 1998, KEK-Tanashi, Tokyo}}

\author{Gino Isidori\thanks{electronic address:
isidori@lnf.infn.it}
\\
{\it INFN, Laboratori Nazionali di Frascati}\\ {\it P.O. Box 13, 00044
Frascati (Rome), Italy } }
\date{}

\begin{document}
\maketitle

\begin{abstract}
A short overview of FCNC and $CP$--violating observables in $K$ decays is
presented. Particular attention is paid to the possibility of performing
precision tests of flavour dynamics and to the search for New Physics.
\end{abstract}

\newpage
\section{Introduction}
The study of kaon decays has historically provided one of the richest
source of information in the construction of the Standard Model (SM). Above
all, let's recall the discovery of $P$ and $CP$ violation, as well as the
indirect indication of the existence of charm. Moreover, at present some of
the most stringent constraints which any extension of the SM has to face on
flavour mixing, $CP$ violation and $CPT$ conservation are derived from kaon
physics. But what is even more fascinating is the fact that in the near
future, 50 years after their discovery, kaon decays could still offer a
valuable and {\em unique} probe to test the SM and to search for New
Physics (NP) \cite{reviews}.

In general, we can separate in three wide classes the
observables which it is still very important to measure with increasing
accuracy: \begin{enumerate}
\item
{\em Pure NP searches.}
The observables belonging to this
class are those vanishing or extremely small within the SM, like the widths
of the lepton--flavour violating modes ($K_L\to \mu e$, $K\to \pi \mu e$,
\ldots) or the transverse muon polarization in $K^+\to \pi^0\mu^+\nu_\mu$
(see e.g. Rizzo in \cite{reviews} and references therein).
The first ones
are completely forbidden within the SM whereas the latter is expected to be
much smaller than the experimental sensitivity. In these cases a
non--vanishing experimental evidence would provide a clear signal for
physics beyond the SM, however a positive result is not guaranteed. \item
{\em Precision SM measurements.}
Under this name we group the observables which are completely dominated by
SM contributions but are calculable with high accuracy
in terms of fundamental parameters. An interesting example in this sector
is provided by the $\pi\pi$ scattering lengths, measurable from $K_{l4}$
decays, which can be expressed in terms of the expectation value of the
quark condensate in the chiral limit \cite{bijnens}. Similarly, $K_{l3}$
decays provide precise information about the Cabibbo
angle and quark--mass ratios \cite{bijnens}.
\item
{\em Short--distance observables.}
In this category we finally collect the $CP$--violating and FCNC observables
which are calculable with high accuracy in terms of short--distance
amplitudes, like the widths of $K\to\pi\nu\bar{\nu}$ decays. This group
is probably the most interesting one since it is useful both to test the
flavour structure of the SM and also to search for NP. In the following we
will concentrate only on this sector, trying to emphasize the cleanliness
from long--distance effects and the NP sensitivity of various observables.
\end{enumerate}

\section{FCNC rare decays within the SM} The rare transitions $K\to\pi
\nu\bar{\nu}$, $K\to \ell^+\ell^-$ and $K\to\pi \ell^+\ell^-$ are naturally
good candidates to extract information on the FCNC amplitude $s_L \to d_L
f_L \bar{f}_L\ (f=\nu,\ell)$. Within the SM this amplitude is generated
only at the quantum level, through $Z$--penguin and $W$--box diagrams, and
is particularly interesting because of the dominant role played by the
top--quark exchange.
Separating the contributions to the amplitude according to the intermediate
up--type quark running inside the loop, one can write
\beq
\cA(s_L \to d_L f_L \bar{f}_L) = \sum_{q=u,c,t} V_{qs}^*V_{qd}~\cA_q~,
\label{uno}
\eeq
where $V_{ij}$ denote the elements of the
Cabibbo--Kobayashi--Maskawa (CKM) matrix \cite{CKM}. The hierarchy of the
CKM matrix \cite{Wolf} would favor the first two terms in (\ref{uno})
however the hard GIM mechanism of the parton--level calculation implies
$\cA_q \sim m^2_q/M_W^2$, which leads to a completely different scenario:
assuming the standard phase convention ($\Im V_{us}=\Im V_{ud}=0$) and
expanding the CKM elements in powers of the Cabibbo angle ($\lambda=0.22$)
\cite{Wolf}, one finds
\beq
V_{qs}^*V_{qd}~\cA_q~\sim~\left\{ \begin{array}{ll} \cO(\lambda^5
m_t^2)~+i~\cO(\lambda^5 m_t^2)\quad     & (q=t)~, \\
\cO(\lambda m_c^2 )~+i~\cO(\lambda^5 m_c^2)     & (q=c)~, \\
\cO(\lambda \Lambda^2_{QCD})    & (q=u)~.
\end{array} \right.
\label{due}
\eeq
As can be noticed, the top--quark contribution dominates both real and
imaginary parts of the amplitude (the $\Lambda^2_{QCD}$ factor in the last
line follows from a naive estimate of long--distance effects associated to
the up--quark exchange). This implies several interesting consequences for
$\cA(s_L \to d_L f_L \bar{f}_L)$: i) it is dominated by short--distance
dynamics and therefore calculable with high precision in perturbation
theory; ii) it is very sensitive to $V_{td}$, which is one of the less
constrained CKM matrix elements;
iii) it is likely to have a large $CP$--violating phase; iv) it is very
suppressed within the SM and thus very sensitive to possible NP effects.

The short--distance contributions to
$\cA(s_L \to d_L f_L \bar{f}_L)$ can be
efficiently described by means of a single effective dimension--6 operator:
$O^{f}_{LL}= (\bar{s}_L\gamma^\mu d_L)(\bar{f}_L \gamma_\mu f_L)$. The
Wilson coefficients of $O^{f}_{LL}$ have been calculated by Buchalla and
Buras including next--to--leading--order QCD corrections \cite{BB} (see
also \cite{MU,BB2}), leading to a very precise description of the partonic
amplitude.
Moreover, the simple structure of $O^{f}_{LL}$ has two major advantages: i)
the relation between partonic and hadronic amplitudes in the above
mentioned rare decays is quite accurate, since the hadronic matrix elements
of the $(\bar{s}_L \gamma^\mu d_L)$ current between a kaon and a pion (or
the vacuum) are related by isospin symmetry to those entering $K_{l3}$ (or
$K_{l2}$) decays, which are experimentally well known; ii) the lepton pair
is produced in a state of definite $CP$ and angular momentum ($J^{CP}=1^-$)
implying, for instance, that the leading contribution of $\cA(s_L \to d_L
f_L \bar{f}_L)$ to $K_L \to \pi^0 f \bar{f}$ is $CP$
violating.

The short--distance contribution of the $s_L \to d_L f_L \bar{f}_L$ amplitude
to $K\to\pi \nu\bar{\nu}$, $K\to \ell^+\ell^-$ and $K\to\pi \ell^+\ell^-$
is therefore very well under control. 
The remaining question to address in order to quantify 
their potential in testing flavour dynamics is the estimate of
other possible contributions. For instance in the case of 
$K\to\pi \ell^+\ell^-$ an important role is certainly 
played by the $s_L \to d_L \ell_V \bar{\ell}_V$ amplitude, due 
to electromagnetic interactions. Then in all decays there is 
the question of possible long--distance contaminations. 
In the following we shall discuss in more detail the potential
sources of uncertainties for the various channels.

\subsection{$K\to\pi \nu\bar{\nu}$}
These modes are particularly clean since neutrinos couple to quarks only
via $W$ and $Z$ exchange, thus the only non--vanishing contribution to the
decay is provided by the $s_L \to d_L \nu_L \bar{\nu}_L$ amplitude
discussed above.

In the charged channel ($K^+\to\pi^+ \nu\bar{\nu}$)
the dominant theoretical error is related to the uncertainty of the QCD
corrections to $\cA_c$
(see \cite{BB2} for an updated discussion), which can be translated into a
$5\%$ error in the determination
of $|V_{td}|$ from $B(K^+\to\pi^+ \nu\bar{\nu})$. This QCD uncertainty can
be considered
as generated by `intermediate--distance' dynamics; genuine long--distance
effects associated to $\cA_u$ have been shown to be substantially smaller
\cite{LW}.

The case of $K_L\to\pi^0 \nu\bar{\nu}$ is even more clean from the
theoretical point of view \cite{Litt}. Indeed, because of the $CP$
structure, the leading contribution to the decay amplitude generated by
dimension--6 operators is proportional to the imaginary parts in
(\ref{uno}). This implies that in the dominant (direct--$CP$--violating)
part of the amplitude the charm contribution is completely negligible with
respect to the top one, where the uncertainty of the QCD corrections is
around 1\%. Intermediate and long--distance contributions to this decay
are essentially confined only to the indirect--$CP$--violating contribution
($K_L \to K_S \to \pi^0 \nu\bar{\nu}$ \cite{BB3}) and to the
$CP$--conserving one (generated at short distances by higher--dimensional
operators \cite{CPC}) which are both extremely small.
Taking into account also the isospin--breaking corrections to the hadronic
matrix elements \cite{MP}, one can therefore write a very accurate
expression (with a theoretical error around $1\%$)
for $B(K_L\to\pi^0 \nu\bar{\nu})$ in terms of short--distance parameters
\cite{BB2,BB3}: \beq
B(K_L\to\pi^0 \nu\bar{\nu})_{SM}~=~4.25 \times 10^{-10}~\left[
\frac{\overline{m}_t(m_t) }{ 170~{\rm GeV}} \right]^{2.3} ~\left[ \frac{\Im
\lambda_t }{ \lambda^5 } \right]^2~. \eeq

The high accuracy of the theoretical predictions of $B(K^+ \to\pi^+
\nu\bar{\nu})$ and $B(K_L \to\pi^0 \nu\bar{\nu})$ in terms of the modulus
and the imaginary part of $\lambda_t= V^*_{ts} V_{td}$ could clearly offer
the possibility of very interesting tests of the CKM mechanism. Indeed, a
measurement of both channels would provide two independent information on
the unitarity triangle (or equivalently on the $\rho$--$\eta$ plane
\cite{Wolf}), which can be probed also by $B$--physics observables. In
particular, as emphasized in \cite{BB2}, the ratio of the two branching
ratios could be translated into a determination of $\sin(2\beta)$, the
$CP$--violating observable measurable in a clean way also from
$B^0(\bar{B}^0)\to J/\Psi K_S$. A comparison of the two measurements would
then provide a very powerful tool to search for NP.

Taking into account all the indirect constraints on $V^*_{ts}$ and $V_{td}$
obtained within the SM, the present range of the SM predictions for the two
branching ratios is given by \cite{BB2}: \beqa
B(K^+ \to\pi^+ \nu\bar{\nu})_{SM} &=& (0.82 \pm 0.32) \times 10^{-10}~,
\label{BRK+nnt}\\
B(K_L \to\pi^0 \nu\bar{\nu})_{SM} &=& (3.1 \pm 1.3) \times 10^{-11}~,
\label{BRKLnnt}
\eeqa
to be compared with the recent experimental results: \beqa
B(K^+ \to\pi^+ \nu\bar{\nu}) &=& 4.2^{+9.7}_{-3.5}\times
10^{-10}~\protect\cite{E787}~, \label{BRK+nnexp} \\ B(K_L \to\pi^0 \nu
\bar{\nu}) &<& 1.6 \times
10^{-6}~\protect\cite{KTEV}~. \label{BRKLnnexp} \label{BRKL}
\eeqa

\subsection{$K\to \ell^+\ell^-$ and $K\to\pi \ell^+\ell^-$} In the decays
involving charged leptons the problem of long--distance effects becomes
much more important because of the presence of electromagnetic
interactions. In general we can distinguish three classes of
electromagnetic long--distance amplitudes: \begin{enumerate}
\item
{\em One--photon exchange.}
This mechanism provides the by far dominant contribution to the
$CP$--allowed transitions $K^+ \to \pi^+ \ell^+ \ell^-$ and $K_S \to \pi^0
\ell^+ \ell^-$ \cite{EPR} (see \cite{DEIP} for an updated discussion). The
former has been observed, both in the electron and in the muon mode,
whereas only an upper bound of about $10^{-6}$
exists on $B(K_S\to\pi^0 e^+e^-)$ \cite{PDG}. Unfortunately chiral symmetry
alone does not help to relate $B(K^+ \to \pi^+ \ell^+ \ell^-)$ and $B(K_S
\to \pi^0 e^+ e^-)$, and without
model--dependent assumptions one can only set a theoretical upper bound of
about $10^{-8}$ on the latter \cite{DEIP}.

In the case of $K_L \to \pi^0 \ell^+ \ell^-$ the long--distance part of the
one--photon exchange amplitude is forbidden by $CP$ invariance but it
contributes to the decay via $K_L$--$K_S$ mixing, leading to
\beq
B(K_L \to \pi^0 e^+ e^-)_{CPV-ind}~=~ 3\times 10^{-3}~ B(K_S \to \pi^0 e^+
e^-)~.
\eeq
On the other hand, the direct--$CP$--violating part of the decay amplitude
is very similar to the one of $K_L \to \pi^0 \nu \bar{\nu}$ but for the
fact that it receives an additional short--distance contribution by the
photon penguin. This theoretically clean part of the amplitude leads to
\cite{BLMM}
\beq
B(K_L\to\pi^0 e^+e^-)^{SM}_{CPV-dir}~=~0.69 \times 10^{-10}~\left[
\frac{\overline{m}_t(m_t) }{ 170~{\rm GeV}} \right]^{2} ~\left[ \frac{\Im
\lambda_t }{ \lambda^5 } \right]^2~. \eeq
The two $CP$--violating components of the $K_L\to\pi^0 e^+e^-$ amplitude
will in general interfere. Given the present uncertainty on $B(K_S \to
\pi^0 e^+ e^-)$, at the moment we can only set the rough upper limit
\beq
B(K_L\to\pi^0 e^+e^-)_{CPV-tot}^{SM}~\lsim~{\rm few}\times 10^{-11}
\label{BRKLet}
\eeq
on the sum of all the $CP$--violating contributions to this mode (the
present experimental upper bound is about two orders of magnitude larger
\cite{PDG}). We stress, however, that the phases of the two $CP$--violating
amplitudes are well know. Thus if $B(K_S \to \pi^0 e^+ e^-)$ will be
measured, it will be possible to determine the interference between direct
and indirect $CP$--violating components of $B(K_L\to\pi^0 e^+e^-)_{CPV}$ up
to a sign ambiguity.

\item
{\em Two--photon exchange in $S$ wave.} This amplitude plays an important
role in $K_L\to\ell^+\ell^-$ transitions. In the $K_L\to e^+ e^-$ case it
is by far the dominant contribution
and it can be estimated with a relatively good accuracy in terms of
$\Gamma(K_L\to \gamma \gamma)$. This leads to the prediction
$B(K_L \to e^+ e^-) \sim 9 \times 10^{-12}$ \cite{VP} which recently seems
to have been confirmed by the four $K_L\to e^+ e^-$ events observed at
BNL--E871 \cite{E871}.

More interesting from the short--distance point of view is the case of $K_L
\to \mu^+\mu^-$. Here the two--photon long--distance amplitude is still
large but the short--distance one, generated by the real part of $\cA(s_L
\to d_L \mu_L \bar{\mu}_L)$ and thus sensitive to $\Re V_{td}$ \cite{BB},
is comparable in size. Unfortunately the dispersive part of the two--photon
contribution is much more difficult to be estimated in this case, due to
the stronger sensitivity to the $K_L \to \gamma^* \gamma^*$ form factor.
Despite the precise experimental determination of $B(K_L \to \mu^+\mu^-)$,
the present constraints on $\Re V_{td}$ from this observable are not very
interesting \cite{DIP}. Nonetheless, the measurement of $B(K_L \to
\mu^+\mu^-)$ is still useful to put stringent bounds on possible NP
contributions. Moreover, we stress that the uncertainty of the $K_L \to
\gamma^*\gamma^*\to \mu^+\mu^-$ amplitude could be partially decreased in
the future by precise experimental information on the form factors of $K_L
\to \gamma \ell^+\ell^-$ and
$K_L \to e^+e^- \mu^+\mu^-$ decays, especially if these would be consistent
with the parameterization of the $K_L \to \gamma^* \gamma^*$ form factor
proposed in \cite{DIP}.

\item
{\em Two--photon exchange in $D$ wave.}
This final amplitude (the smallest of the three) is interesting since it
produces a
non--helicity--suppressed $CP$--conserving contribution to $K_L \to\pi^0 e^+
e^-$ \cite{KLggee}. This contribution does not interfere with the
$CP$--violating one in the total rate and leads to
$B(K_L \to\pi^0 e^+ e^-)_{CPC}~\sim~{\rm few}\times 10^{-12}$. At the
moment it is not easy to perform accurate predictions of $B(K_L \to\pi^0
e^+ e^-)_{CPC}$, however, precise experimental information on the
di--photon spectrum of $K_L\to\pi^0\gamma\gamma$ at low $m_{\gamma\gamma}$
could help to clarify the situation \cite{KLggee}. Moreover, the Dalitz
plot distribution of $CPV$ and $CPC$ contributions to $K_L \to\pi^0 e^+
e^-$ are substantially different: in the first case the $e^+e^-$ pair is
in a $P$ wave, whereas in the latter it is in a $D$ wave. Thus in principle it
is possible to
extract experimentally the interesting
$B(K_L \to\pi^0 e^+ e^-)_{CPV}$ from an observation of various $K_L
\to\pi^0 e^+ e^-$ events. \end{enumerate}

\section{$K\to \pi \nu \bar{\nu}$ and $K_L \to\pi^0 e^+ e^-$
beyond the SM}

As we have seen in the previous section, the branching ratios of $K_L \to
\pi^0 \nu \bar{\nu}$, $K^+ \to \pi^+ \nu \bar{\nu}$ and $K_L \to\pi^0 e^+
e^-$\footnote{~The measurement of $B(K_L \to\pi^0 e^+ e^-)$ should be
supplemented by a Dalitz plot analysis and a determination
or a stringent experimental bound on $B(K_S \to\pi^0 e^+ e^-)$.} could give
us valuable and precise information about flavour mixing. Within the SM
this is ruled by the CKM mechanism, which implies the strong
$\cO(\lambda^5)$ suppression of $\cA(s_L \to d_L f_L \bar{f}_L)$ and leads
to the small predictions in (\ref{BRK+nnt}--\ref{BRKLnnt}) and
(\ref{BRKLet}).
It is therefore natural to expect that these observables are very
sensitive to possible extensions of the SM in the flavour sector.

As long as we are interested only in NP effects to rare FCNC processes, we
can roughly distinguish the extensions of the SM into two big groups: those
involving new sources of flavour mixing (like generic SUSY extensions of the
SM, models with new generations of quarks, etc\ldots) and those where the
flavour mixing
is still ruled by the CKM matrix (like the 2--Higgs--doublet model of type
II, constrained SUSY models, etc\ldots). In the second case the effect to 
rare decays is typically small, at most of the same order of magnitude
as the SM contribution (see e.g. \cite{WWZ,SUSYc} for some recent
discussions). On the other hand, in the
first case it is easy to generate sizable effects,
leading to large enhancements with respect to the SM rates (see e.g.
\cite{CI} and \cite{fourth}).

Interestingly, despite the variety of NP models, it is possible to derive a
model--independent relation among the widths of the three neutrino modes
\cite{GN}. Indeed, the isospin structure of any $s\to d$ operator bilinear
in the quark fields implies
\beq
\Gamma(K^+\to\pi^+\nu\bar{\nu}) =
\Gamma(K_L\to\pi^0\nu\bar{\nu}) +
\Gamma(K_S\to\pi^0\nu\bar{\nu}) ~,
\label{Tri}
\eeq
up to small isospin--breaking corrections, which then leads to
\beq
B(K_L\to\pi^0\nu\bar{\nu})~ <~ \frac{\tau_{_{K_L}}}{\tau_{_{K^+}}}
B(K^+\to\pi^+\nu\bar{\nu})~.
\label{GNbd}
\eeq
Any experimental limit on $B(K_L\to\pi^0\nu\bar{\nu})$ below this bound can
be translated into a non--trivial dynamical information on the structure of
the $s\to d\nu\bar{\nu}$ amplitude. Using the experimental result in
(\ref{BRK+nnexp}), the present model--independent bound on
$B(K_L\to\pi^0\nu\bar{\nu})$ is about $6 \times 10^{-9}$ (more than two
orders of magnitude larger than the SM value!).

Unfortunately there is no analog model--independent bound for $K_L \to\pi^0
e^+ e^-$. However, to compare the NP sensitivity of $K_L \to\pi^0 \nu
\bar{\nu}$ and $K_L \to\pi^0 e^+ e^-$, we note that in the specific
scenario where the dominant contribution to both processes is generated by
an effective $Z{\bar s}d$ vertex, one expects
$B(K_L\to\pi^0  e^+ e^- )\simeq B(K_L \to\pi^0\nu\bar{\nu} )/6$ \cite{CI}.

\subsection{Supersymmetric contributions}
We will now discuss in more detail the rare FCNC transitions
in the framework of a low--energy supersymmetric extension of the SM
--with unbroken $R$ parity, minimal particle content and generic flavour
couplings--
which represents a very attractive possibility from the theoretical
point of view.  Similarly to the SM, also in this case FCNC
amplitudes are generated only at the quantum level.
However, in addition to the standard penguin and box diagrams, also their
corresponding superpartners,  generated by gaugino--squarks loops, play an
important role. In particular, the chargino--up--squarks diagrams provide
the potentially dominant non--SM effect to the $s \to d \nu \bar{\nu}
(\ell^+\ell^-)$ amplitude \cite{MWBRS}. Moreover, in the limit
where the average mass of SUSY particles ($M_S$)
is substantially larger than $M_W$,
the penguin diagrams tend to dominate over the box ones and the dominant
SUSY effect can be encoded through an effective
$Z \bar{s}_L d_L$ coupling \cite{CI}.

The flavour structure of a generic SUSY model is quite complicated
and a convenient way to parametrize the various flavour---mixing
terms is provided by the so--called mass--insertion approximation
\cite{HKR}. This consists of choosing a simple basis for the gauge 
interactions and, in that basis, to perform a 
perturbative expansion of the squark mass matrices
around their diagonal. The same approach could be employed also within the
SM, rotating for instance the $u_L^i$ fields and choosing the basis
where the $W-d_L-u_L^d$ coupling is diagonal. 
 In this case it would be easy to verify that the dominant contribution
to the $Z \bar{s}_L d_L$ vertex is generated at the second order in 
the mass expansion by a double $q^i_L - q^j_R$ mixing, namely $(u^{d}_L - t_R)
\times (t_R - u^{s}_L)$. The two off--diagonal mass terms would 
indeed be proportional to  $m_t V_{td}$ and $m_t V^*_{ts}$. 
As shown in \cite{CI}, this ``second--order structure''
remains valid also for the SUSY (chargino--up--squarks)
contributions. In this case 
the situation is slightly more complicated due to the
interplay between the standard CKM matrix (ruling the higgsino--$q^i_L -
\tq^j_R$ vertex) and a new matrix responsible for the $\tq^i_L - \tq^j_R$
mixing \cite{MWBRS}. It is indeed possible to consider terms with a
single off--diagonal CKM element and a single $\tq^i_L - \tq^j_R$ mixing.
However, in perfect analogy with the SM case, the potentially dominant
SUSY contribution arises from the double mixing $(\tiu^{d}_L - \tit_R) \times
(\tit_R - \tiu^{s}_L)$ \cite{CI}. This  leads to an effective 
$Z \bar{s}_L d_L$ vertex proportional to 
\beq
{\tilde \lambda}_t=  
 \frac{ ({\tM}^2_U)_{s_L t_R} ({\tM}^2_U)_{t_R d_L} }{ M_S^4}~, 
\label{eq:ldef}
\eeq
which can be considered as the analog of the SM factor
$\lambda_t (m_t^2/M_W^2)$. 

The phenomenological constraints 
on ${\tilde \lambda}_t$  can be divided into two groups: 
\begin{enumerate}
\item     
indirect $M_S$--dependent bounds on 
$({\tM}^2_U)_{s_L t_R}$ and $({\tM}^2_U)_{t_R d_L}$, dictated 
mainly by vacuum--stability, neutral--meson mixing 
($K^0-\bar{K}^0$, $D^0-\bar{D}^0$ and $B^0-\bar{B}^0$)
and $b \to s \gamma$;  
\item  
direct limits on the $Z \bar{s}_L d_L$ 
coupling dictated by $K_L\to\mu^+\mu^-$ and $\Re(\epoe)$,
constraining $\Re {\tilde \lambda}_t$ and $\Im {\tilde \lambda}_t$,
respectively.
\end{enumerate}
In a wide range of $M_S$ (0.5 TeV $\lsim M_S\lsim$ 1 TeV) 
the first type of 
bounds are rather weak and leave open the possibility 
for large effects in rare decays. In particular, 
$\Gamma(K^+ \rightarrow \pi^+ \nu \bar\nu)$ could be enhanced up 
to one order of magnitude with respect to the SM prediction,
whereas for $\Gamma(K_L \rightarrow \pi^0 \nu \bar\nu)$
and $\Gamma(K_L \rightarrow \pi^0 e^+ e^-)$ the enhancement 
could even be higher \cite{CI}. 
Concerning the direct constraints, 
the bound on $\Re {\tilde \lambda}_t$ from $K_L\to\mu^+\mu^-$
is certainly quite stringent \cite{BS}, 
however one could still generate the above large 
enhancements with an almost imaginary ${\tilde \lambda}_t$
(actually this is a necessary condition to enhance 
the $CP$--violating modes). 

Buras and Silvestrini recently claimed that
the possibility of a large $\Im {\tilde \lambda}_t$  is 
substantially reduced by the constraints from $\Re(\epoe)$ 
\cite{BS}. According to these authors, the enhancement of 
the rare widths can be at most of one order of magnitude 
in $\Gamma(K_L \rightarrow \pi^0 e^+ e^-)$ and 
not more than a factor $\sim 3$ in 
$\Gamma(K^+ \rightarrow \pi^+ \nu \bar\nu)$.
We agree with them that in principle the measurement 
of $\Re(\epoe)$ provides a bound on $\Im {\tilde \lambda}_t$,
however we are more skeptical about the precise value of 
this bound at present. 
As we shall discuss more extensively in the next section,  
the problem with $\Re(\epoe)$ is that on one side 
the SM prediction is affected by large theoretical uncertainties,
on the other side this observable is sensitive also to other SUSY
effects, which could partially cancel the contribution of 
$\Im {\tilde \lambda}_t$.
In addition, even the experimental results 
concerning $\Re(\epoe)$ are not very clear at present \cite{PDG}.
Probably the situation will improve in the future, but 
at the moment the extraction of bounds on 
the $Z \bar{s}_L d_L$ vertex from  $\Re(\epoe)$ 
requires some additional assumptions. 
On the contrary, we stress that the direct constraints which could be
obtained from the rare decays, even if less stringent,  
would be much more clear from the theoretical point of view.

\section{$\epoe$ within and beyond the SM}
The $\epoe$ parameter can be  defined as
\beq
\frac{\epsilon'}{\epsilon}~ = ~ 
\frac{ e^{ i({\pi/2}+\delta_2-\delta_0)}}{\epsilon} 
\frac{\omega }{ \sqrt{2}} \left[
{\Im A_2 \over \Re A_2}-{\Im A_0 \over \Re A_0}\right]~,
\eeq
where $A_{0,2}$ denote the $K^0\to (2\pi)_{0,2}$ amplitudes, 
$\delta_{0,2}$ the corresponding strong phases,
$\omega=\Re A_2/\Re A_0 \simeq 1/22$ and $\epsilon$ is 
the standard $\Delta S=2$ $CP$--violating term. 
A measurement of $\epoe$ can provide very interesting 
information about the global symmetries of the SM. Indeed,
as it is well known, an evidence for $\epoe\not=0$
would be a clear signal of direct $CP$ violation 
\cite{Maiani}. Moreover, given that 
arg$(\epsilon) = \pi/4 \simeq {\pi/2}+\delta_2-\delta_0$,
the phase of $\epoe$ is almost vanishing, implying
$|\Im (\epoe)| \ll |\Re (\epoe)|$. This relation can 
be modified only by adding $CPT$ non--invariant terms
in $K\to 2\pi$ amplitudes 
and thus can be used to test $CPT$ invariance \cite{Maiani}.   

More problematic is the question of what kind of short--distance 
information can be extracted from $\epoe$ and thus to what extent 
this observable can be used to perform precision tests of the
SM in the flavour sector. Similarly to the rare $FCNC$ 
transitions, also the weak phases of $A_0$ and $A_2$
are generated only at the quantum level and 
are very sensitive to the structure of the CKM matrix.
The short--distance information about these amplitudes 
are usually encoded in the Wilson coefficients of
appropriate four--quark operators, which can
be calculated with a good accuracy down to scales
$\mu\gsim m_c$  \cite{Munich,Rome}.
However, contrary to the rare decays, in the case of 
$K\to 2 \pi$ transitions it is very difficult to 
evaluate the hadronic matrix elements of the 
effective operators.

At the quark level $\Im A_0$ is dominated by 
the gluon penguin whereas $\Im A_2$ 
by the electroweak ones. In both cases the dominant 
contribution is provided by four--quark operators 
of the type $(\bar{s}^\alpha_L\gamma^\mu d^\beta_L) \sum_q y_q
(\bar{q}^\beta_R \gamma_\mu q^\alpha_R)$, 
namely $O_6$  for $\Im A_0$ ($y_q=1$) and  $O_8$
for $\Im A_2$  ($y_q=e_q$),  
which have enhanced matrix elements in the chiral limit. 
A useful approximate expression for $\Re(\epoe)$ can be
obtained by showing explicitly the dependence 
on the matrix elements of these two operators \cite{BS,epeBu}:
\beq
\Re \left(\frac{\epsilon'}{\epsilon}\right)_{SM}~ = ~ \left[-1.4
+8.2 \left(R_s B_6^{(1/2)}\right) - 4.0 \left(R_s B_8^{(3/2)}\right) 
\right]\times \Im \lambda_t~.
\label{eqdue}
\eeq
Here $R_s=[ 158~{\rm MeV}/(m_s(m_c)+m_d(m_c)) ]^2$ 
shows the leading dependence on the quark masses of the 
two matrix elements, whereas their actual value is 
hidden in the $B$--factors
$B_6^{(1/2)}$ and $B_8^{(3/2)}$, expected to
be positive and $\cO(1)$. The uncertainty 
in the numerical coefficients of (\ref{eqdue}) is expected
to be around or below $20\%$ \cite{epeBu} (see also 
Buras in \cite{reviews}). 

Various estimates of $R_s$ and of 
the $B$--factors can be found in the literature, 
leading to results for $\Re(\epoe)_{SM}$ which range  essentially
between $0$ and $3\times 10^{-3}$ \cite{epeBu,epeCi,epeBe}. 
Certainly some non--perturbative techniques are more reliable 
than others, however in all cases it is very difficult to 
provide quantitative estimates of the errors, especially in the case 
of the $B$--factors.
Lattice results, for example, are based on the lowest--order 
chiral relation between $\langle K|O_i| 2\pi\rangle$ 
and $\langle K|O_i| \pi\rangle$, and could be affected by
sizable corrections due to
next--to--leading terms in the chiral expansion.
Interesting progress in calculating hadronic matrix elements
have recently  been made in the framework of the 
$1/N_c$ expansion \cite{hambye,deRafael}, nonetheless 
even there we are still far from precise results, especially in the 
case of $O_6$ and $O_8$. 

Given the above considerations, it is clear that at present 
$\Re(\epoe)$ cannot be used to perform
precision tests of the SM. In the context of NP scenarios, 
one can generally expect two main  effects 
in $\Re(\epoe)$: i) a modification of the phase of the
gluon--penguin amplitude and thus of $\Im A_0$, 
ii) a modification of the phase of the electroweak--penguin 
amplitude and thus of $\Im A_2$. As we have shown in the 
previous section, the second effect could be bounded 
independently also from the rare processes 
$K_L \to \pi^0 \nu\bar{\nu}$ and $K_L \to \pi^0 e^+ e^+$.  
In the future $\Re(\epoe)$ could therefore provide an 
interesting complementary window for NP searches in $\Delta S=1$ 
amplitudes. However, this would require better experimental bounds on
both rare modes and $\Re(\epoe)$ and, possibly, also better
theoretical control on the $B$--factors.

\section{Conclusions}
The $K \to \pi \nu\bar{\nu}$ decays provide a unique opportunity to
perform high precision tests of $CP$ violation and flavour mixing,
both within and beyond the SM. In some NP scenarios, even 
in the case of generic supersymmetric extensions of the SM, 
sizable enhancements to  $B(K \to \pi \nu\bar{\nu})$ are 
possible and, if detected, these could provide the first evidence for physics 
beyond the SM. However, even if NP will be discovered 
before via direct searches, we stress that precise measurements 
of these rare modes will provide unique information about the 
flavour structure of any extension of the SM. 

Among the $K \to X_d \ell^+\ell^-$ decays, the most
interesting one from the short--distance point of view is
probably $K_L \to \pi^0 e^+ e^-$.  In order to extract 
precise information from this mode, the measurement of 
its decay rate should be accompanied by a Dalitz plot analysis and 
a determination or a stringent experimental bound 
on $B(K_S \to\pi^0 e^+ e^-)$. 

Accurate measurements of $\Re(\epoe)$ and $\Im(\epoe)$
will provide interesting information about the global 
symmetries of the SM (especially if $\Re(\epoe)$ were
found to be clearly different form zero). However, given the large 
theoretical uncertainty, at present $\Re(\epoe)$ is not very 
useful to perform precision tests of the model.

\section*{Acknowledgments}
It is a pleasure to thank Shojiro Sugimoto, Taku Yamanaka 
and the other organizers of the ``International Workshop
on $CP$ Violation in $K$'' for the hospitality in Tokyo 
and for providing such a pleasant and stimulating atmosphere. 
I am grateful also to Gerhard Buchalla and Gilberto Colangelo for 
the recent fruitful collaborations on $K\to\pi\nu\bar{\nu}$ decays
and for interesting comments. Finally, I wish to thank Andrzej  
Buras and Luca Silvestrini for useful discussions.


\end{document}